\input{psfig.sty}

\documentstyle[twoside,11pt]{article}

\newcommand{\mathPic}[2]{
\begin{figure*}[bt]
  \centerline{\psfig{figure=figs/#1.ps,width=4.5in,height=3in,bbllx=0.5in,bblly=2.75in,bburx=8in,bbury=8in}}
\label{#1}
\caption{#2}
\vspace{0.33in}
\end{figure*}
}

\newcommand{\qed}{\hfill\noindent\rule[.2em]{.7em}{.7em}}

\newcommand{\proof}{\noindent{\em Proof.}\ }

\newcommand{\Choose}[2]{{#1 \choose #2}}

\newtheorem{lemma}{Lemma}
\newtheorem{corollary}{Corollary}

\renewcommand{\baselinestretch}{1}  

\begin{document}

\title{\Large \bf
Faster Parametric Shortest Path and \\ Minimum Balance Algorithms}

\author{
\bf Neal E. Young\thanks{Research supported by the Hertz Foundation.}  \\
  \sl Computer Science Department, Princeton University, Princeton,  \\
  New Jersey, 08544 \\
\and
\bf Robert E. Tarjan\thanks{Research at Princeton University partially
     supported by National Science Foundation Grant DCR-8605961 and
     Office of Naval Research Contract N00014-87-K-0467.} \\
  \sl Computer Science Department, Princeton University, Princeton,  \\
  NJ 08544, and NEC Research Institute, Princeton, NJ 08540. \\
\and
\bf James B. Orlin\thanks{Research partially supported by NSF PYI
    grant 8451517-ECS, AFOSR grant AFOSR-88-0088, and grants from Analog
    Devices, Apple Computers, Inc., and Prime Computer.}  \\
  \sl Sloan School of Management, Massachusetts Institute of Technology,  \\
  Cambridge, MA 02139. \\
}
\date{}

\maketitle

\footnotetext{
\vspace{0.2in}
\begin{flushleft}
NETWORKS, Vol. 21:2 (March, 1991) \\
\copyright 1991 by John Wiley \& Sons, Inc.
\end{flushleft}
\vspace{-0.8in}
}

\vspace{0.5in}
{ \small\baselineskip=9pt
We use Fibonacci heaps to improve a parametric shortest path algorithm
of Karp and Orlin, and we combine our algorithm and the method of
Schneider and Schneider's minimum-balance algorithm to obtain a faster
minimum-balance algorithm.

\baselineskip=9pt
For a graph with $n$ vertices and $m$ edges, our parametric shortest
path algorithm and our minimum-balance algorithm both run in 
$O(nm + n^2\log n)$ time, improved from $O(nm\log n)$ for the parametric
shortest path algorithm of Karp and Orlin and $O(n^2 m)$ for the
minimum-balance algorithm of Schneider and Schneider.

\baselineskip=9pt
An important application of the parametric shortest path algorithm is
in finding a minimum mean cycle.  Experiments on random graphs suggest
that the expected time for finding a minimum mean cycle with our
algorithm is $O(n\log n + m)$.
}
\vspace{0.2in}

\section{\normalsize \bf INTRODUCTION}

The body of the paper contains five sections.  The first section
describes the parametric shortest path problem and an algorithm for
solving it that runs in $O(nm + n^2\log n)$-time on an $n$-vertex
graph with $m$ edges.  The algorithm is based on an $O(nm\log n)$-time
algorithm of Karp and Orlin \cite{KO-81}, modified to take
advantage of the Fibonacci heap data structure of Fredman and Tarjan
\cite{FT-87}. 
The second section describes the minimum mean cycle problem and how
the parametric shortest path algorithm can be used to solve it.  

The third section describes the minimum-balance problem and an
algorithm for solving it that runs in $O(nm + n^2\log n)$-time.  The
algorithm combines the method of Schneider and Schneider
\cite{SS-87}, which yields a straightforward $O(n^2 m)$-time
algorithm for the problem, with the parametric shortest path
algorithm.

The fourth section describes the results of implementing the
parametric shortest path algorithm for finding a minimum mean cycle
and running it on random graphs.  The results suggest that the
expected time for the parametric shortest path algorithm to find a
minimum mean cycle is close to $O(m+n\log n)$, and that even for small
graphs the algorithm is faster than the $O(nm)$-time algorithm of Karp
\cite{karp-78}.  

A solution to the parametric shortest path problem is given by a
sequence of trees, which our algorithm generates but does not store.
The final section discusses how the trees may be implicitly stored so
that any tree in the sequence can be generated quickly.  Also
considered in the final section are generalizations of the problems
to which our algorithms still apply.

\section{\normalsize \bf PARAMETRIC SHORTEST PATHS}
\label{KO81Sec}

The parametric shortest path problem is a generalization of the
standard single-source shortest path problem in which some of the edge
costs have a parameter subtracted from them.  An instance of the
problem is specified by giving a directed graph $G=(V,E,c)$ with edge
costs, a source vertex $s$ with all vertices reachable from $s$, and a
subset $E'$ of the edges representing those edges whose costs have the
parameter subtracted from them.  Specifically, a particular value
$\lambda$ of the parameter yields the directed graph $G_\lambda=(V,E,c
- \lambda\delta_{E'})$ with edge costs, where
$(c-\lambda\delta_{E'})(e) = c(e) - \lambda\delta_{E'}(e)$, and
$\delta_{E'}(e) = 1$ if $e\in E'$ and $0$ otherwise.  (We adopt the
convention that the parameter is subtracted so that influence of
parameterized edges on shortest paths increases with the parameter.)

The problem is to determine a shortest path tree in $G_\lambda$ for
every $\lambda$ such that shortest paths in $G_\lambda$ are well
defined.  It is well known that shortest paths in $G_\lambda$ are well
defined if and only if $G_\lambda$ contains no negative-cost cycle.
Thus the problem is to determine a shortest path tree for each
$G_\lambda$ such that $\lambda\in[-\infty,\lambda^*]$, where
$\lambda^*$ is as large as possible such that $G_{\lambda^*}$ has no
negative-cost cycle.  If $G_\lambda$ has no negative-cost cycle for
all $\lambda$, then we take $\lambda^* = \infty$.  In $G_\infty$, we
take shortest paths to be those that are shortest in $G$ among those
that have the maximum number of parameterized edges.  Similarly, if
$G_\lambda$ has a negative-cost cycle for all $\lambda$, we take
$\lambda^* = -\infty$, and take shortest paths in $G_{-\infty}$ to be
those that are shortest in $G$ among those which have the minimum
number of parameterized edges.

A solution to the problem is given by a finite sequence of trees
$T_0,T_1,\ldots,T_k$ and a finite non-decreasing sequence of real
numbers
\(-\infty = \lambda_0\le\lambda_1\le\cdots\le\lambda_{k+1} = \lambda^*\)
such that $T_i$ is a shortest path tree in $G_\lambda$ for all
$\lambda$ in $[\lambda_i,\lambda_{i+1}]$.  A solution could also be
given by a sequence of trees and {\em strictly} increasing real
numbers.  The algorithm we give may produce sequences with some
$\lambda_i$ equal to $\lambda_{i+1}$.  If the second type of solution
is desired, such $\lambda_i$ and the corresponding $T_i$ can simply be
removed from the sequence.

Applications of the parametric shortest path problem include the
minimum concave-cost dynamic network flow problem \cite{GO-85}, matrix
scaling \cite{OR-85,SS-89}, and the minimum mean cycle and minimum
balancing problems, discussed below.

\subsection{\normalsize \bf An Inductive Method}

A natural method for solving the parametric shortest path problem is
to proceed tree by tree.  That is, determine $T_0$ and then
inductively determine successive $\lambda_i$ and $T_i$.  This is the
method that we use.

The first tree $T_0$ for $\lambda_0 = -\infty$ can be determined by
finding a shortest path tree from $s$ by running a standard
$O(nm)$-time single source shortest path algorithm on $G_\alpha$,
where $\alpha < -\sum_{e\in E}|c(e)|$.  For this value of the
parameter, paths with fewer parameterized edges always cost less than
paths with more, so a shortest path tree in $G_\alpha$ is also a
shortest path tree in $G_{-\infty}$.  The shortest path algorithm that
is used must be able to detect the case when a negative-cost cycle
exists (shortest paths are not well defined), for this will be the
case if $\lambda^* = -\infty$.

\subsection{\normalsize \bf Pivot Paths}

Next we consider the induction step.  Suppose that tree $T$ is a
shortest path tree from $s$ in $G_\lambda$.  Consider increasing the
parameter from $\lambda$ until it reaches a value $\lambda'$ beyond
which $T$ ceases to be a shortest path tree.  The reason that $T$
ceases to be a shortest path tree is that some path $p$ not in $T$
from $s$ to some vertex $v$ becomes shorter than its counterpart $t_v$
(the path from $s$ to $v$) in $T$.  In order for this to happen, $p$
must be equal in cost to $t_v$ in $G_{\lambda'}$ and have more
parameterized edges than does $t_v$.  We call $\lambda'$ the {\em pivot
point from $T$}, and any such path $p$, a {\em pivot path for $T$}.

How can we find a pivot path --- a shortest path in $G_{\lambda'}$
with more parameterized edges than has the corresponding path in $T$ ---
without knowing $\lambda'$?  One way would be to consider each path
$p$ from $s$ with more parameterized edges than has its corresponding path
in $T$.  Of these paths, if we choose one that will first become equal
in cost to its corresponding path in $T$ as the parameter is
increased, we will have a pivot path.  

Once we know a pivot path $p$ for $T$, we can determine the pivot
point $\lambda'$, because the costs of $p$ and $t_v$ can coincide at
only one value of the parameter.  In $G_{\lambda'}$, $p$ and its
counterpart $t_v$ in $T$ are both shortest paths, and thus they and
their corresponding prefixes are all of equal cost.  Thus all paths in
$T\cup p$ are shortest paths.  If we construct the subgraph $T'$ by
deleting all edges of $T$ that lead into vertices of $p$ and adding
the edges of $p$, then, provided $T'$ is a tree, it is a shortest path tree
in $G_{\lambda'}$.

To make sure we are making progress in discovering the subgraph $T'$,
we will rule out {\em degenerate} pivot paths --- those with a proper
prefix with fewer parameterized edges than has the corresponding path in
$T$ or with a zero-cost cycle with no parameterized edges.  From any
degenerate pivot path, we can construct a nondegenerate pivot path by
replacing the offending proper prefix by the corresponding path in
$T$ or by deleting the offending cycle.  For the rest of the paper
when we refer to a pivot path we will mean a nondegenerate pivot
path.

This gives us the following algorithm.  Determine an initial shortest
path tree $T_0$ for $G_{-\infty}$.  Look for a (nondegenerate) pivot
path $p$ for $T_0$.  If there is none, stop.  Otherwise, use this path
to determine the pivot point and a subgraph $T'$.  If $T'$ is not a
tree, stop.  Otherwise, take $T'$ to be the next shortest path tree and
continue.

If the algorithm stops because there are no pivot paths, then the
current shortest path tree will continue to be a shortest path tree
indefinitely as the parameter is increased, and $\lambda^* = \infty$.

Otherwise, the algorithm may stop because $T'$ is not a tree.  The
subgraph $T'$ is formed by replacing the edges in the tree $T$ that
lead into the vertices on the pivot path with the pivot path edges.
Thus, $T'$ can only fail to be a tree if it contains a cycle and can
only contain a cycle if the pivot path $p$ contains a cycle.  But if
$p$, a shortest path, contains a cycle, it must be a zero-cost cycle.
Since $p$ is nondegenerate, the cost of the cycle must be decreasing
with the parameter.  Thus the graph will have a negative-cost cycle
for any larger value of the parameter and $\lambda^* = \lambda'$.

Thus, if the algorithm terminates, it gives a correct sequence of
trees and intervals.  To bound the number of trees produced by the
algorithm, consider the number of parameterized edges on the path into
each vertex in $T$ and $T'$.  By the construction of $T'$ and the
nondegeneracy of the pivot path, the path in $T'$ into a vertex
contains at least as many parameterized edges as does the path in $T$.
Furthermore, at least one vertex in $T'$ has more parameterized edges.
Thus, the sum over all vertices of the number of parameterized edges on
the path into the vertex in the current tree increases by at least one
with each new tree.  Since this sum is nonnegative and bounded above
by $n^2$ (actually $n(n-1)/2$), it follows that the algorithm
produces at most $n^2$ trees.

\subsection{\normalsize \bf A Faster Implementation}

First, we will reduce the number of potential pivot paths that the
algorithm checks.  Suppose there is a (nondegenerate) pivot path, and
let $p$ denote its shortest prefix that is still a pivot path.  Let
$v$ denote the destination vertex of $p$.  Any proper prefix of $p$ is
not a pivot path, yet is a shortest path in $G_{\lambda'}$ and has at
least as many parameterized edges as does the corresponding path in $T$.
Thus, each proper prefix of $p$ has the same number of parameterized
edges as does its counterpart in $T$, and if we replace any proper prefix
of $p$ by its counterpart in $T$, we obtain a pivot path.  In
particular, if we replace the largest proper prefix of $p$ by its
counterpart in $T$, we obtain a pivot path that consists of a path in
$T$ followed by an edge (necessarily not in $T$).  For each edge
$e=(u,v)$, we let $p(e)$ denote the path in $T$ from $s$ to $u$
followed by the edge $e$.  Thus, if any pivot path exists, some $p(e)$
is a pivot path.  We call any such $p(e)$ a {\em canonical pivot
path}.

In order to find a canonical pivot path quickly, we will associate
with each edge $e=(u,v)$ the value of the parameter, if any, at which
the cost of the path $p(e)$ will become equal to the cost of $t_v$.
Since $t_v$ is of cost no more than that of $p(e)$ for the current
value of the parameter, this value will exist if and only if $p(e)$
has more parameterized edges than does $t_v$.  In this case, we call
the value the {\em key} of the edge.  Otherwise, the key of the edge is
taken to be infinity.  More specifically, let $c(p)$ denote the cost
of a path $p$ in $G$ and let $\delta_{E'}(p)$ denote the number of
parameterized edges on $p$, so that $c(p)-\lambda\delta_{E'}(p)$ is
the cost of a path $p$ in $G_{\em \lambda'}$.  Then, the key of $e$ is
\[\frac{c(p(e))-c(t_v)}{\delta_{E'}(p(e)) - \delta_{E'}(t_v)}\]
provided $\delta_{E'}(p(e)) > \delta_{E'}(t_v)$, and infinity
otherwise.

Since $p(e)$ is just $t_u$ followed by $e$, we can calculate edge keys
in constant time if we maintain for each path $t_w$ the values
$c(t_w)$ and $\delta_{E'}(t_w)$.  When a pivot occurs and the tree
changes, we can find (and update) the vertices for which these values
change by searching in $T$ from the end of the pivot path.
Furthermore, these values only change for a vertex when it acquires a
new shortest path, so the time to maintain these values over the
course of the algorithm is proportional to the number of shortest path
changes over the course of the algorithm, and, thus, at most $n^2$.

This gives us an implementation running in $O(n^2 m)$ time --- maintain
the tree $T$, maintain the values $c(t_w)$ and $\delta_{E'}(t_w)$, and perform
each of the at most $n^2$ pivots by choosing the minimum edge key in
$O(m)$ time to define a pivot path.

If we store the edge keys in a standard heap data structure, the time
to find each minimum is reduced to $O(\log n)$, so the total time to
find minimum keys is reduced to $O(n^2\log n)$.  The time to maintain
the keys increases to $O(\log n)$ per key change, but since the key of
an edge is changed only when one of its endpoints acquires a new path,
the total number of edge key changes during the course of the
algorithm is at most $2nm$.  (Recall that each time a vertex acquires
a new shortest path, the number of parameterized edges on its current
path increases, so that no vertex acquires a new path more than $n-1$
times during the course of the algorithm.)  Thus the total time spent
maintaining keys is $O(nm\log n)$.  This implementation of the
algorithm, which is the implementation given by Karp and Orlin
\cite{KO-81}, therefore runs in time $O(nm\log n)$.

\subsection{\normalsize \bf Vertex Keys}

A complete, but different, exposition of the $O(nm\log n)$-time
algorithm described above was given by Karp and Orlin in 1981, before
the discovery by Fredman and Tarjan of the Fibonacci heap data
structure \cite{FT-87} in 1984.  The advantage of the F-heap data
structure is that the time taken to decrease or insert a key is $O(1)$
in the amortized sense \cite{Tar-85}.  The time taken to find the
minimum key or increase a key is $O(\log n)$ in the amortized sense.
Although we can construct graphs yielding $\Theta(nm)$ key increases,
so that storing the edge keys in an F-heap does not immediately give a
faster algorithm, we can still use an F-heap to our advantage.

To do this, we associate with each vertex $v$ a key that is the
minimum of the keys of the edges entering $v$.  This is the value of
the parameter at which the cost of one of the potential pivot paths
$p(e)$ into $v$ will first become equal to the cost of the current
path into $v$.  With this value we associate the edge $e$.

The minimum vertex key still yields a canonical pivot path.  When a
pivot occurs, and a vertex acquires a new shortest path, what effect
does that have on the vertex keys?  Recall that when a vertex acquires
a new shortest path, the new path has more parameterized edges.  The
cost of the new path equals the cost of the old path at the pivot
point, but is decreasing faster with the parameter.  How the key of a
vertex $v$ changes depends on whether the path into $v$ changes and
how the potential pivot paths into $v$ change.  Suppose the vertex
does not acquire a new shortest path and none of the potential pivot
paths into it change.  In this case, the key remains the same.  If the
vertex does not acquire a new shortest path but some potential pivot
path into it changes, then the cost of the new potential pivot path is
decreasing faster than is the cost of the old.  In this case, the new
path will overtake the current path into $v$ sooner, so if the vertex
key changes, it will only decrease.

If a vertex acquires a new shortest path, and none of the potential
pivot paths change, then the cost of the new shortest path, which is
decreasing faster, will not be overtaken as soon as the old.  Thus, in
this case, the key will increase.  For the remaining case, note that
the number of parameterized edges on each new shortest path exceeds
the number on the coresponding old path by the same amount.  Thus, in
this case, the key will stay the same if any of the potential pivot paths that
determined the minimum value previously also change, and, otherwise, the
key will increase.

To summarize, a vertex key is the minimum of the keys of the edges
into the vertex.  The algorithm stores with each vertex key the edge
whose key determines the value of the vertex key.  As before, the
algorithm determines a pivot path from the minimum key, computes the
new tree $T'$, and updates the values $c(t_w)$ and $\delta_{E'}(t_w)$
for each $w$ that acquires a new shortest path.  To maintain the
vertex keys, for each vertex that acquires a new shortest path, the
algorithm examines the keys of the edges coming into the vertex and
takes the new vertex key to be the minimum (possibly increasing the
key) and checks the key of each outgoing edge to see if it has
decreased below the key of the vertex at the other end, and if it has,
it updates that vertex key.  It then continues pivoting, as before.

The purpose of this modification is that now the algorithm needs to do
fewer increase key operations in the worst case, so that the time
taken by the algorithm is reduced by storing the keys in an F-heap.
In particular, we will see next that the time taken to maintain keys,
which dominates the time taken by the algorithm, is reduced from
$O(nm\log n)$ to $O(nm + n^2\log n)$.

\subsection{\normalsize \bf Running Time}

\label{boundSec}

To bound the time taken by the algorithm, we note that the
initialization of the data structures takes $O(m)$-time, plus
$O(nm)$-time if a shortest path algorithm needs to be run to determine
the initial tree.  To bound the remaining time, we will associate each
operation involved in updating the data structures with a shortest
path change to some vertex and then bound the total number of
shortest path changes during the course of the algorithm.  We give a
slightly more detailed analysis than necessary, which will be useful
in Section \ref{expSec}.

Let $m_w$ and $j_w$ denote the degree of vertex $w$ and the number of
shortest path changes (``jumps'') to $w$ in the course of the
algorithm, respectively.  As noted above, for any graph, $j_w \le
n-1$.  Once pivoting begins, finding a pivot path takes amortized time
$O(\log n)$.  At each pivot some vertex changes path, so the time for
finding pivot paths is $O(\log n \sum_w{j_w})$.  After a pivot is
found, for each vertex that changes path, the shortest path information
for the vertex is updated, the edges into and out of the vertex are
examined, the vertex key may be increased, and the vertex keys of
adjacent vertices may be decreased.  Thus, each time $w$ changes path,
the amortized time to maintain the data structures is $O(\log n +
m_w)$.  [Recall that the amortized times for increase key and decrease
key operations in the F-heap are, respectively, $O(\log n)$ and
$O(1)$.]  Thus, the time taken by the algorithm after initialization is
bounded by a constant times
\begin{eqnarray}
& & \sum_w{j_w m_w} + \log n \sum_w{j_w} \label{boundEqn}\\
& \le & n \sum_w{m_w} + \log n \sum_w{j_w} \nonumber \\
& \le & 2 n m + n^2 \log n. \label{worstBound}
\end{eqnarray}
Thus, the algorithm always runs in time $O(nm + n^2 \log n)$.

\section{\normalsize \bf THE MINIMUM MEAN CYCLE PROBLEM}
\label{minMeanCycleSec}

The {\em minimum mean cycle problem} for a graph with cycles is to
find a directed cycle in the graph that minimizes the average cost of
the edges on the cycle.  The average edge cost of such a cycle is
called the {\em minimum cycle mean}.  Solutions to this problem are
needed in a minimum-cost circulation algorithm of Goldberg and Tarjan
\cite{GT-89} and in a graph minimum-balancing algorithm of Schneider
and Schneider \cite{SS-87}.  The problem has been studied by Karp
\cite{karp-78}, who gave an $O(nm)$-time dynamic programming
algorithm, and by Ahuja and Orlin \cite{AO-88-2}, who gave an
$O(\sqrt{n}m\log{nC})$-time scaling algorithm.  (Here, $C$ is the
maximum of the edge costs, which must be integers for the Ahuja-Orlin
algorithm to work correctly.)

As Karp and Orlin \cite{KO-81} have observed, the minimum mean
cycle problem can be solved using an algorithm for the parametric
shortest path problem.  Before we discuss how, we introduce the
concept of a {\em potential} for a graph.  Potentials are related to
the dual variables arising when path and flow problems are formulated
as linear programs.  They are also an inherent part of the minimum
balance problem that we discuss in the next section.

A potential is an assignment of real-valued weights to the vertices of
the graph. Such a potential acts to change the edge costs of the graph,
as follows:  The cost of an edge $e=(u,v)$ has the weight of $u$ added
to it and the weight of $v$ subtracted from it.  Thus, potential
$\pi:V\rightarrow\Re$ acting on the graph $G=(V,E,c)$ produces the
graph $G^\pi=(V,E,c^\pi)$, where $c^\pi(e = (u,v)) = c(e) + \pi(u) -
\pi(v)$.  One useful aspect of potentials is that they do not change
shortest paths or costs of cycles.

It is well known that for any graph with no negative-cost cycle, there
is a potential for which the resulting graph has all nonnegative edge
costs.  Specifically, we fix a source vertex $s$ in the graph $G$ from
which all vertices are reachable.  (If there is no such vertex in the
original graph, we introduce an artificial source vertex with
zero-cost edges to all other vertices.  The minimum mean cycle is
unchanged by this alteration.)  Let $\pi(v)$ be the cost of the
shortest path from $s$ to $v$.  Then, the well-known inequality $\pi(v)
\le \pi(u) + c(e)$ holds for every edge $e=(u,v)$.  The cost of edge
$e$ in $G^\pi$ is therefore nonnegative.  Furthermore, if $e$ lies on
a shortest path, or on a zero-cost cycle, then equality holds in the
above relation, so the transformed edge cost is zero.  We call $\pi$ a
{\em shortest path potential for $G$}.

Using the notation of Section \ref{KO81Sec}, with $E=E'$, let
$G_\lambda$ denote the graph $G$ with $\lambda$ subtracted from all
edge costs, and let $\lambda^*$ denote the largest $\lambda$ such that
shortest paths in $G_\lambda$ are well defined.  As shown earlier, if
$\lambda^*\ne\infty$, then $G_{\lambda^*}$ has a zero-cost cycle, but
no negative-cost cycle.  It follows that if we apply the shortest path
potential $\pi$ for $G_{\lambda^*}$ to $G_{\lambda^*}$, we obtain a
graph with a zero-cost cycle and nonnegative edge costs.  If we then
add $\lambda^*$ to all edge costs, we obtain the graph $G^\pi$.  It
follows that $G^\pi$ has a cycle $C$ with each edge of cost
$\lambda^*$ and that no edge in $G^\pi$ has cost less than
$\lambda^*$.  Thus, $C$ is a minimum mean cycle of mean cost
$\lambda^*$ in $G^\pi$.  Since potentials do not change cycle costs,
$C$ is also a minimum mean cycle in $G$ and the minimum cycle
mean in $G$ is $\lambda^*$.

Thus, to solve the minimum mean cycle problem, it suffices to obtain a
shortest path potential for $G_{\lambda^*}$.  Such a potential is
easily obtainable from the last shortest path tree produced by the
parametric shortest path algorithm run on $G$ with all edges
parameterized.  In this case, the algorithm stops because it discovers
a zero-cost cycle in $G_{\lambda^*}$, which, in turn, is a minimum mean
cycle in $G$.  Thus, the minimum mean cycle problem is easily reduced
to the parametric shortest path problem.

As a side note, the introduction of an artificial source vertex with
zero-cost edges into all the other vertices is useful even if an
original source vertex is available.  An initial shortest path tree
from the artificial source vertex is given by the zero-cost edges, so
there is no need to solve an arbitrary single-source shortest path
problem.  This variant corresponds to the {\em sourceless}
parameterized shortest path problem, where shortest paths independent
of source into each node are computed rather than shortest paths from
a particular source.  This variation also applies to Karp's minimum
mean cycle algorithm \cite{karp-78} and simplifies it by removing
an initial strongly connected components computation.  In practice,
however, partitioning the graph into strongly connected components
might reduce the solution time.

\section{\normalsize \bf THE MINIMUM BALANCE PROBLEM}

We say that a subset of the vertices of $G$ is {\em minimum-balanced}
if the minimum cost among edges entering the subset is the same as is
the minimum cost among edges leaving the subset.  The {\em
minimum-balance} problem for a strongly connected graph $G=(V,E,c)$ is
to determine a potential $\pi$ such that, in $G^\pi$, each subset of
the vertices other than $E$ or $\emptyset$ is minimum-balanced.  Such
a potential is said to {\em minimum-balance} $G$.  Schneider and
Schneider introduced an equivalent problem \cite{SS-87} in connection
with matrix balancing, and they gave an $O(n^2m)$-time algorithm.  Our
algorithm, which runs in time $O(nm+n^2\log n)$, can be viewed as a
faster implementation of Schneider and Schneider's algorithm.

To solve the minimum balance problem, note that finding a shortest
path potential $\pi$ for $G_{\lambda^*}$ and a minimum mean cycle $C$
is a step toward minimum-balancing the graph.  In $G^\pi$, each subset
of vertices that $C$ enters and leaves has edges entering and leaving
of cost exactly $\lambda^*$, and no entering or leaving edge has cost
less than $\lambda^*$.  Consider contracting $C$ to a single new
vertex $v$ in $G^\pi$, deleting self-loops but retaining multiple
edges, to obtain the graph $H$. Let $\beta$ be a potential that
minimum-balances $H$.  If $H$ has only one vertex, $\beta$ can be
taken to be any function; otherwise it can be obtained recursively.
The edges of $H^\beta$ are all of cost no less then the minimum cycle
mean of $H$, which is, in turn, no less than $\lambda^*$, the minimum
cycle mean of $G$.

To complete the minimum-balancing, we can essentially just add the
potentials $\pi$ and $\beta$.  Consider extending the potential
$\beta$ for $H$ to the potential $\alpha$ for $G^\pi$ defined by
$$\alpha(w) = \cases{\beta(v)&$w$ on $C$;\cr \beta(w)&
otherwise.\cr}$$ The effect of $\alpha$ on $G^\pi$ is as follows.
Each edge in $G^\pi$ that corresponds to an edge in $H$ becomes of
cost equal to the cost of the corresponding edge in $H^\beta$.  The
other edges, joining vertices on the cycle $C$, remain unchanged in
cost.  Thus, if a subset of the vertices is entered and left by $C$, it
is entered and left by edges of cost $\lambda^*$, which are minimum.
Otherwise, the edges entering and leaving the subset correspond to the
edges entering and leaving the corresponding subset of the vertices of
$H$, and so the subset is correspondingly minimum-balanced.  Thus, the
potential $\pi+\alpha$ minimum-balances $G$.

In summary, the minimum-balancing algorithm repeatedly finds and contracts
a minimum mean cycle until the graph contains only one vertex.  Each
time the cycle is contracted, a shortest path potential is computed
for the current graph.  The minimum-balancing potential is computed by
adding together the shortest path potentials, appropriately extended.

At this point we have reduced the minimum-balance problem to a series
of at most $n$ problems involving finding minimum mean cycles and
shortest path potentials.  Schneider and Schneider, who gave the above
method in \cite{SS-87}, also noted that Karp's $O(nm)$-time algorithm
for finding minimum mean cycles can be extended to yield shortest path
potentials, thus obtaining an $O(n^2m)$-time algorithm.

\subsection{\normalsize \bf A Hybrid Algorithm}

The successive graphs computed by the above method are closely
related.  By modifying the parametric path algorithm to contract the
minimum mean cycle it discovers and continue, we obtain a faster
algorithm.

Suppose that we run the parametric shortest path algorithm on $G$
with an arbitrary source $s$ and $E=E'$.  Since $G$ is strongly
connected, the algorithm will terminate, having produced $\lambda^*$ and
a tree $T_k$ that is a shortest path tree in $G_{\lambda^*}$.  This is
sufficient to obtain a shortest path potential for $G_{\lambda^*}$,
as required by the first iteration of the minimum-balance algorithm.

The minimum-balance algorithm would next adjust the edge costs of $G$ by
applying the shortest path potential $\pi$, contract the resulting
graph around the discovered minimum mean cycle, and continue.
Consider what happens if instead of stopping the parametric shortest
path algorithm at this point we try to continue it through the
adjustment and contraction.

First, to what extent can we maintain the shortest path tree?  We have
a tree $T$ that is a shortest path tree in $G_{\lambda^*}$.  Since
potentials preserve shortest paths, $T$ is also a shortest path tree
in $G'=G_{\lambda^*}^\pi$.  We would like a shortest path tree in
$G'/C$, $G'$ contracted around $C$.  By properties of the shortest
path potential, we know that edge costs in $G'$ are nonnegative and
that edges on $C$ or $T$ are of zero cost in $G'$.  It follows that
$T/C$, the tree $T$ contracted around the cycle $C$, has all zero-cost
edges in $G'/C$.  Since $C$ consists of a single edge not in $T$
together with a path in $T$, $T/C$ is a tree in $G'/C$.  Thus, $T/C$ is
a shortest path tree in $G'/C$.

Once the shortest path tree $T/C$ is constructed, we can completely
recompute the secondary data structures, including $c(t_w)$ and
$\pi_{E'}(t_w)$ for each vertex $w$, the F-heap of vertex keys, and
the mapping of which edges have determined which vertex keys, in
$O(m)$-time.  We are then ready to proceed with the inductive step of
the parametric shortest path algorithm, modifying the tree pivot by
pivot until the next minimum mean cycle is found.

\subsection{\normalsize \bf Termination and Running Time}

The algorithm continues pivoting and contracting, calculating the
requisite shortest path potentials at each contraction, until the
graph contains only a single vertex.  Note that the initial graph and
thus all subsequent graphs are strongly connected.  Thus, every graph
with at least two vertices will have a cycle with a parameterized
edge, and so the only way a sequence of pivots can stop is by
discovery of a minimum mean cycle.

Since at most $n$ contractions can take place, the time spent by the
algorithm performing contractions and reinitializing data structures
following a contraction is $O(mn)$.  The remaining time is spent
pivoting from one tree to the next between contractions.  The analysis
of the parametric path algorithm bounding the number of pivots and the
time spent maintaining the data structures after each pivot in terms
of the number of path changes continues to apply here.  Each pivot
still results in some vertex acquiring a new path, and each operation
maintaining the data structures is associated with a path change for
some vertex.  To obtain the same worst-case bound (\ref{worstBound}), 
it suffices to note that the number of shortest path changes during
the course of the algorithm is still bounded by $n$ per vertex.  

Consider, as the minimum-balance algorithm proceeds, the number of
vertices in the graph minus the number of parameterized edges on the
shortest path to a vertex $w$.  Every time the path changes as the
result of a pivot, this quantity decreases, and contraction does not
increase it.  Furthermore, the quantity initially is at most $n$ and
on termination is nonnegative.  Thus, it is decreased at most $n$
times, and the number of path changes associated with $w$ is at most
$n$.

\section{\normalsize \bf EXPECTED RUNNING TIMES}
\label{expSec}

Although there are graphs for which the worst-case bounds for the
parametric shortest path and minimum-balance algorithms are tight, one
might suspect that for many, if not most, graphs the bounds are not
tight.  In the case of the minimum balance algorithm, it may be that,
for most graphs, most vertices do not acquire $\Omega(n)$ new shortest
paths throughout the course of the algorithm, either because when a
vertex acquires a new shortest path that path tends to be
substantially longer than the old one, or because the contracted
cycles tend to be larger than constant size.  Since the work done in the
parametric shortest path algorithm is essentially the work done in the
minimum balance algorithm before the first contraction, one might
expect that the work done by the parametric shortest path algorithm
would be even less.

To explore this, consider the behavior of the parametric path
algorithm as used for finding a minimum mean cycle on random directed
graphs.  We can rewrite the bound (\ref{boundEqn}) as follows:
\begin{eqnarray}
& & \sum_w{j_w m_w} + \log n \sum_w{j_w} \nonumber \\
& \le & \max_w{m_w} \sum_w{j_w} + \log n \sum_w{j_w} \nonumber \\
& = & \sum_w{j_w}\left( \max_w{m_w} + \log n\right). \label{expBound}
\end{eqnarray}
With high probability, the maximum degree of a random directed graph
is $O(m/n + \log n)$, so that it remains only to estimate the
number of path changes when the algorithm is run on random directed
graphs.

\mathPic{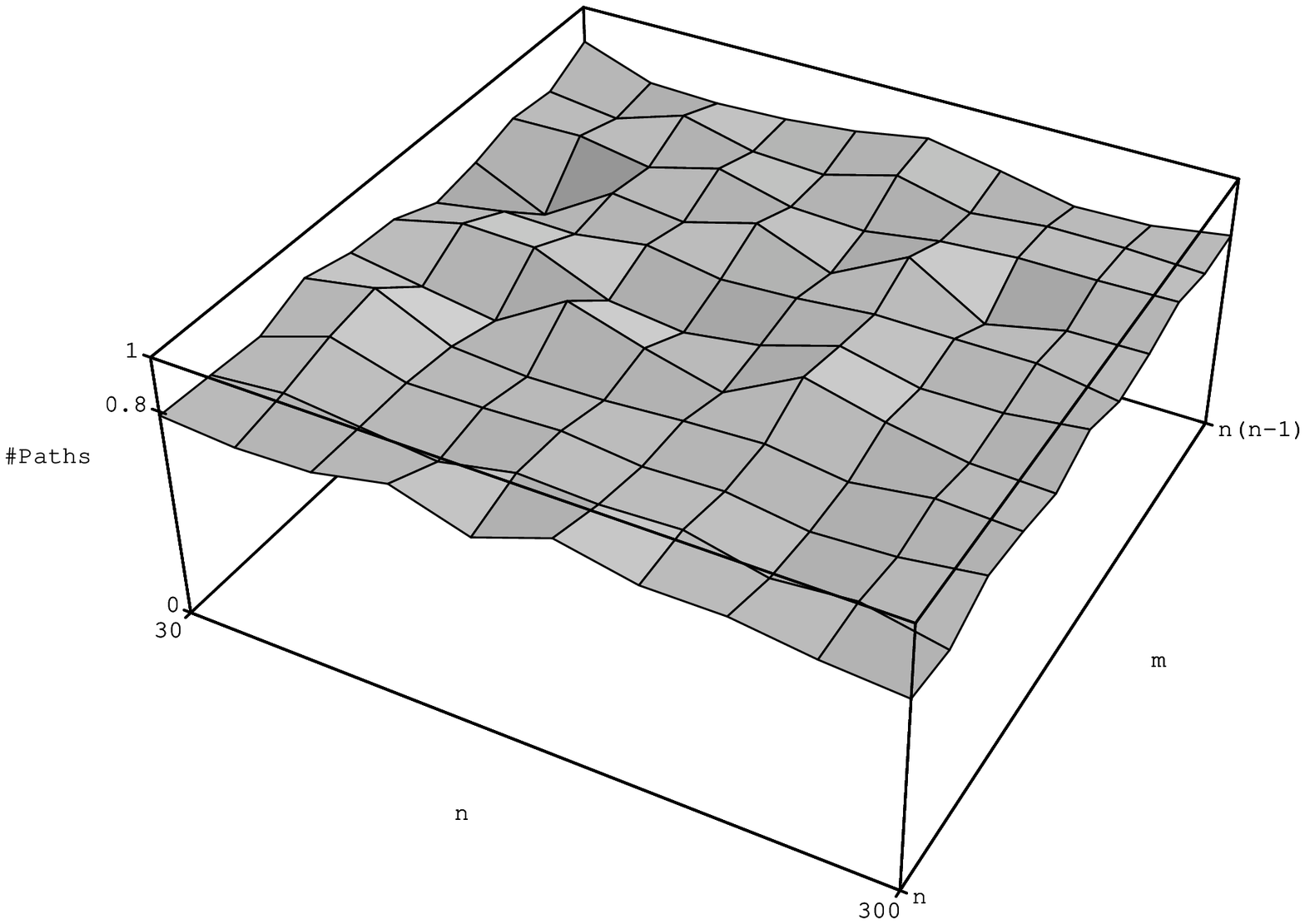}{Average number of shortest path changes
per vertex when finding a min. mean cycle.}

To do this, we ran the sourceless variant of the algorithm on random
directed graphs of $n$ nodes and $m$ edges with each of the $n(n-1)$
edges equally likely to be present.  For each $n$ and $m$, we took
$n/2$ or fifty trials, whichever was larger, and averaged the
number of path changes in each trial.  Figure \ref{jumpFig} shows the
average number of path changes per vertex for the parametric path
algorithm for finding a minimum mean cycle.

The results suggest that the expected number of path changes for the
parametric path algorithm is $O(n)$.  If this is true, then the
parametric path algorithm yields a minimum mean cycle algorithm with
worst case time $O(nm+n^2 \log n)$ and expected time $O(m + n \log n)$.

We make this argument precise in the following lemma and corollary.

\begin{lemma}
\label{degreeLemma}
Given a random directed graph with $n$ nodes and $m$ edges, the
probability that some vertex is of degree higher than
$8m/n+2\log n$ is less than $2/n$.
\end{lemma}
\proof
If $8m/n \ge 2(n-1)$ or $n \le 8$, then clearly the lemma holds.
Assume that $m < \frac{1}{4}n(n-1)$ and $n > 8$.
The number of graphs with $n$ nodes and $m$ edges
with a given vertex of degree $d$ is
\begin{eqnarray*}
g_d & = & \Choose{2(n-1)}{d}\Choose{n(n-1)-2(n-1)}{m-d} \\
 & = & g_{d-1}\frac{(2(n-1)-d+1)(m-d+1)}{d\left(n(n-1)-2(n-1)-(m-d)\right)} \\
 & \le & g_{d-1} \frac{2(n-1)m}{d((n-2)(n-1)-(m-d))} \\
 & \le & g_{d-1} \frac{2(n-1)m}{d((n-2)(n-1)-\frac{1}{4}n(n-1))} \\
 & = & g_{d-1} \frac{8m}{d(3n-8)} \\
 & \le & g_{d-1} \frac{4m}{dn}.
\end{eqnarray*}
Thus, letting $D = \left\lceil 8m/n \right\rceil$,
for $d > D$, 
$g_d \le \frac{1}{2} g_{d-1} \le g_D 2^{-(d-D)}$.
Therefore. the number of graphs with a given vertex of degree greater than
or equal to $D + 2 \log n$ is bounded by 
\[\sum_{d \ge D+2\log n} g_D 2^{-(d-D)}
= g_D \sum_{i \ge 2\log n} 2^{-i} 
\le 2 g_D n^{-2}.\] 
Thus the probability that a given vertex is of degree $D+2\log n$ or more is
less than $2/n^2$.  It follows that the the probability that
some vertex is this degree or higher is less than $2/n$.
\qed

\begin{corollary}
\label{finalCorollary}
If the expected number of path changes is $O(n)$, then the expected
running time of the algorithm is $O(m + n\log n)$.
\end{corollary}
\proof
Let $R$ be the running time of the algorithm on a random graph, let
$B$ be the quantity $\sum_w{j_w}\left( \max_w{m_w} + \log n\right)$,
let $D$ be the event that all vertices of the graph have degree less
than $8m/n+2\log n$, and let $J$ be the number of path changes.
Then, bounds (\ref{worstBound}) and (\ref{expBound}) and lemma
\ref{degreeLemma} give
\begin{eqnarray*}
E(R) & \le & E(B) \\
& = & \Pr[D]E(B|D) + \Pr\left[\bar{D}\right]E(B|\bar{D}) \\
& \le & E(J\times \left(8\frac{m}{n} + 3\log n\right) | D) \\
& &  \ +\frac{2}{n}O(nm+n^2\log n) \\
& = & O(m + n\log n) \left(\frac{E(J | D)}{n}+1\right) \\
& \le & O(m + n\log n) \left(\frac{E(J)}{\Pr[D]n} + 1\right) \\
& = & O(m + n\log n).
\end{eqnarray*} 
\qed

\mathPic{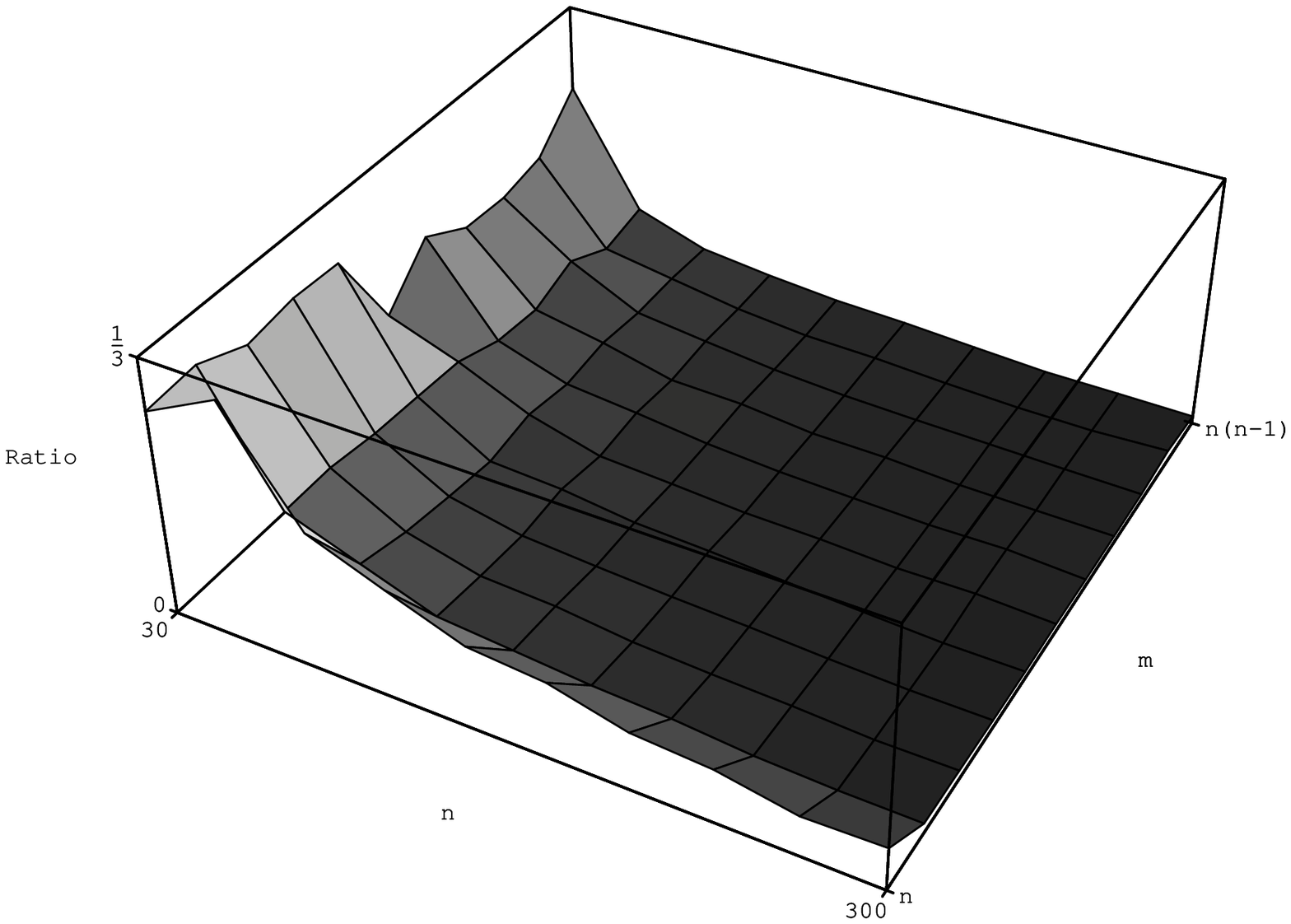}{Ratio of average time to average time for Karp's algorithm.}

Figure \ref{ratioFig} shows the ratio of the average time for
our minimum mean cycle algorithm to the average time of Karp's
$O(nm)$-time minimum mean cycle algorithm, which runs in time
$\Theta(nm)$ for all graphs.  

Although we did not implement the scaling minimum mean cycle algorithm
of Ahuja and Orlin \cite{AO-88-2}, experience with a related
algorithm suggests that even if the scaling algorithm runs in expected
time $O(n\log n+m)$ for some reasonable distribution of graphs, the
constants involved would be larger than those observed for our
algorithm \cite{Sch-89}.

\mathPic{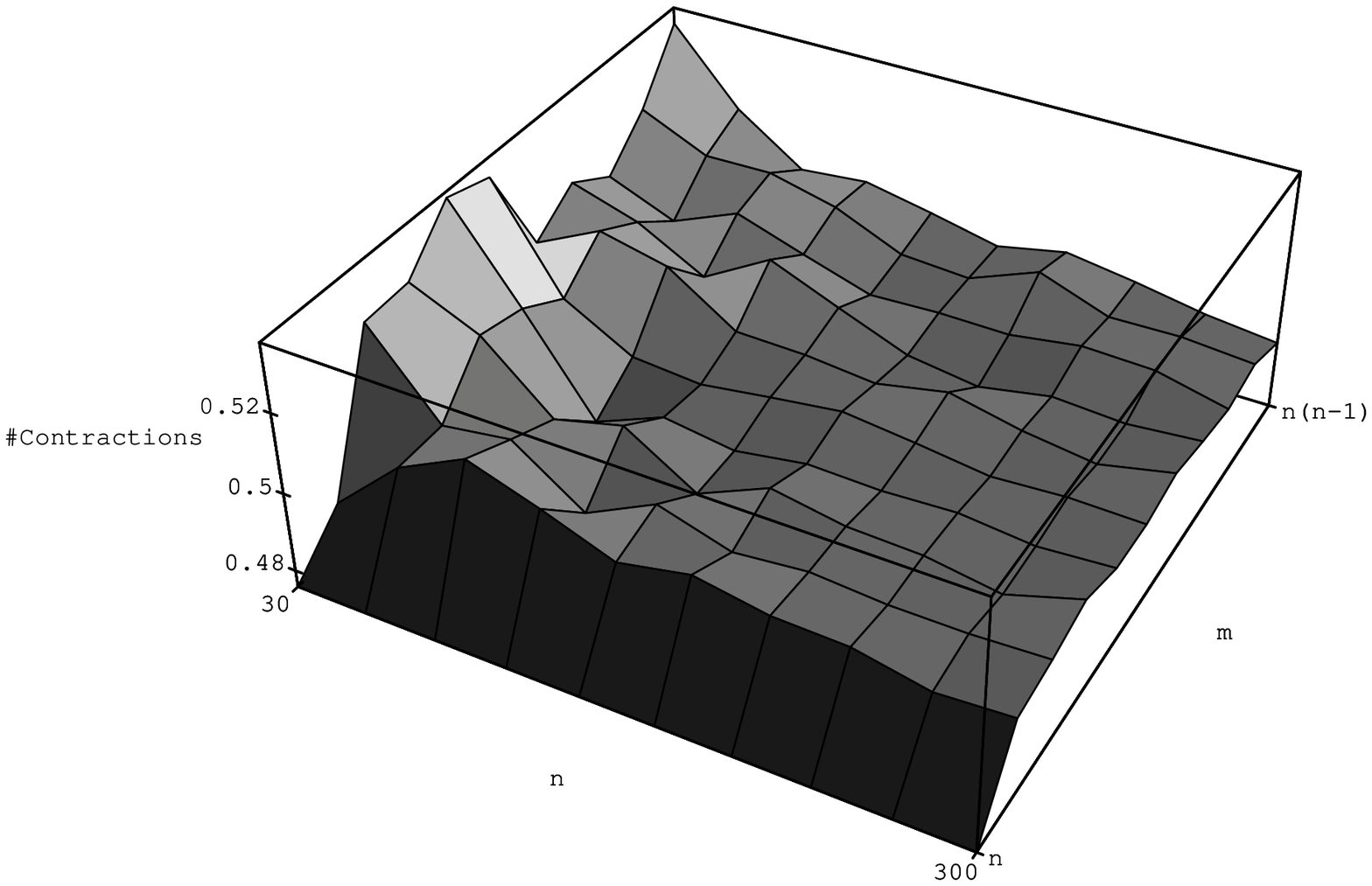}{Average number of contractions per vertex.}

We also tested the minimum balance algorithm.  As Figure
\ref{contractFig} shows, it appears that the number of contractions
for a nonsparse random graph is about $n/2$.  In the algorithm as
described, the contraction step, which takes $O(m)$-time per
contraction, appears to be the bottleneck, giving a running time of
$\Theta(nm)$.  The total time for contraction can be reduced to
$O(J+n\log n)$, where $J$ is the number of shortest path changes, by
using a variant of the union-find data structure of Tarjan
\cite{Tar-79}.  The time for adding the partial potentials is
similarly reduced, so this modification should remove the bottleneck.
We have not estimated the expected running time of the modified
algorithm.

\section{\normalsize \bf FINAL REMARKS}

One question not addressed by our parametric path algorithm is the
form in which the solution is produced.  Recording the sequence of
trees explicitly would take space and time $\Theta(n^3)$ in the worst
case of $\Theta(n^2)$ trees, and recovering a tree for a particular
value of the parameter would require locating the value in the
sequence of intervals, which can be done easily in $\log n$ time.
Alternatively, we can store for each vertex the values of the
parameter at which the parent changes and which vertex becomes the
parent at each change.  Then, in the worst case, the space is $O(n^2)$
and any tree can be recovered in $O(n\log n)$-time by searching for
the parent of each node individually with a binary search.  For most
graphs, the number of times each vertex changes parent is probably
constant, so this is probably an even better solution in practice.

On a different note, a number of generalizations of our algorithms are
possible.  The parametric path algorithm can be generalized to handle
the case in which the edge costs are more general functions of the
parameter.  In particular, if the edge costs are concave functions of
the parameter with derivative in the set $\{-K,\ldots,-1,0\}$, then
essentially the same algorithm works, provided that the functions are
stored so that for each function we can tell for successive values of
the parameter what the current value and derivative are and when the
next decrease in derivative will occur.  For instance, if we allow the
initial graph to contain multiedges, and each multiedge is given as
a list of edges with cost functions of constant derivative in
$\{-K,\ldots,1,0\}$, in order of decreasing derivative, then pivots
still occur when the shortest path into a vertex changes and each
such change decreases the derivative of the cost of the path into the
vertex.  Essentially the same analysis applies to show that the number
of path changes to each vertex is in this case at most $Kn$ and the
algorithm produces at most $Kn^2$ trees in time $O(Knm+Kn^2\log n)$.

Surprisingly, the parametric path algorithm can also be generalized to
allow concave edge cost functions with positive derivative.
If the functions are concave functions of the parameter with
derivative in the range $[-K,..,K]$, then the same argument shows that
the same type of transition from one shortest path tree to the
next occurs.  In this case, finding an initial shortest path tree
is more difficult, however, because an initial value of the parameter
for which shortest paths are well defined is not so easy to come by.

If all edge cost functions are nonnegative at some point (say zero),
then we can start by finding a shortest path tree in $G_0$ and proceed
by increasing the parameter, generating trees in sequence as before,
until some cycle becomes of zero cost.  Then, we can return to $G_0$
and proceed by {\em decreasing} the value of the parameter to generate
the initial part of the sequence of trees in reverse.  We leave open
the problem of finding an initial value of the parameter in the
general case.

The {\em minimum average weighted length cycle problem} \cite{GM-84} is
a generalization of the minimum mean cycle problem in which each edge
of the graph has a (positive) weight as well as a cost.  The problem
is to find the cycle that minimizes the ratio of the sum of the costs
of the edges on the cycle to the sum of weights of the edges on the
cycle.  This problem can be reduced to the generalized parametric
shortest path problem described above, just as the minimum mean cycle
problem can be reduced to the parametric shortest path problem, by
taking the cost function of an edge to be the parameter times the
weight minus the cost.

We also might consider generalizing the minimum-balance problem
by allowing the algorithm to proceed with more general parameterizations.
The analysis of the running time still holds, but in this case we know
of no natural interpretation of the problem that the algorithm is solving.

\bibliographystyle{abbrv}
\small
\bibliography{minBal}

\normalsize

\vspace{0.6in}

Received January, 1990

Accepted April, 1990





\end{document}